\documentclass[reprint,groupedaddress]{revtex4-1}%,linenumbers,showpacs,groupedaddress
\usepackage{amsmath,amssymb,mathrsfs,amsfonts}
\usepackage{txfonts}%txfonts, fourier, mathdesign, pxfonts, arev, cmbright, mathptmx, mathpazo % font style
\usepackage{upgreek}
\usepackage{bm}
\usepackage{color}
\usepackage{appendix}
\usepackage{graphicx}
\usepackage{footmisc}
\usepackage{bm}
\usepackage{color}
\usepackage{CJKutf8}
\usepackage{epsfig}
 \usepackage{float}
\usepackage[dvipdfm,colorlinks,
            citecolor=blue,
            anchorcolor=red,
            menucolor=red,
            linkcolor=blue,
            filecolor=red,
            runcolor=red,
            frenchlinks=red]{hyperref}

\begin{document}
%\captionsetup[figure]{labelfont={bf},labelformat={default},labelsep=period,name={Figure.}}

\begin{CJK}{UTF8}{gbsn}
  \title{Magnetically tunable exciton valley coherence in monolayer WS$_2$ mediated by the electron-hole exchange and exciton-phonon interactions}

   \author{Kang Lan$^1$}
   \author{Shijie Xie$^1$}
   \email{xsj@sdu.edu.cn}
   \author{Jiyong Fu$^2$}
   \email{yongjf@qfnu.edu.cn}
%\email{Corresponding author.\\ yongjf@qfnu.edu.cn}
  \author{Fanyao Qu$^3$}
\affiliation{$^1$ School of Physics, State Key Laboratory of Crystal Materials, Shandong University, Jinan 250100, China\\
$^2$ Department of Physics, Qufu Normal University, Qufu 273165, China\\
$^3$ Instituto de F\'{\i}sica, Universidade de Bras\'{\i}lia, Bras\'{\i}lia-DF 70919-970, Brazil}

\begin{abstract}
    We develop a model, which incorporates both intra- and intervalley scatterings to master equation, to explore exciton valley coherence in monolayer WS$_2$ subjected to magnetic field.
    For linearly polarized (LP) excitation accompanied with an \emph{initial} coherence, our determined valley dynamics manifests the coherence decay being faster than the exciton population relaxation,  and agrees with experimental data by Hao \emph{et al.}[\href{https://dx.doi.org/10.1038/nphys3674}{Nat. Phys. 12, 677 (2016)}].
    Further, we reveal that magnetic field may quench the electron-hole (e-h) exchange induced pure dephasing---a crucial decoherence source---as a result of lifting of valley degeneracy, allowing to \emph{magnetically} regulate valley coherence.
    In particular, at low temperatures for which the exciton-phonon (ex-ph) interaction is weak, we find that the coherence time is expected to attain $\uptau_{\mathcal{C}}\sim 1$ ps, facilitating full control of qubits based on the valley pseudospin.
    For dark excitons, we demonstrate an emerging coherence even in the absence of initial coherent state, which has a long coherence time ($\sim 15$ ps) at low temperature.
    Our work provides an insight into \emph{tunable} valley coherence and coherent valley control based on dark excitons.
\end{abstract}

\maketitle

%\section{introduction}
%\label{sec:intr}
\textit{Introduction}---
Among emerging new two-dimensional materials utilized for valleytronic applications~\cite{doi:10.1073/pnas.0502848102}, monolayer transition metal dichalcogenides (TMDCs) MX$_2$ (M=Mo, W; X=S, Se, Te) have attracted intense research interest, following the discovery of direct band gap at the two inequivalent $K$ and $K^\prime$ valleys of the Brillouin zone~\cite{PhysRevLett.108.196802}.
Owing to the reduced dielectric screening and large electron (and hole) effective masses, the light-matter interaction in TMDCs is dominated by tightly Coulomb-bound excitons with the binding energy up to hundreds of meV~\cite{PhysRevB.93.205423,PhysRevLett.113.026803}.
Also, the space inversion asymmetry together with spin-orbit interaction endows TMDCs the spin-valley \emph{locked} band structure~\cite{PhysRevLett.108.196802,PhysRevLett.119.137401,xu_yao2014}, which enables optical generation and manipulation of valley polarization as well as extraction of valley pseudospin information with the aid of circularly polarized (CP) light~\cite{PhysRevB.90.155449,ugeda2014,yecao2014,PhysRevLett.113.076802,PhysRevB.97.115425,doi:10.1063/1.5112823}, and facilitates control of valley coherence of excitons.

Benefiting from the valley selective transition rule~\cite{PhysRevLett.108.196802}, the initial coherent state for excitons is typically
generated \emph{via} optical pumping of linearly polarized (LP) light~\cite{PhysRevLett.115.117401,jones_yu2013,PhysRevLett.123.096803,2019,hao_moody2016,Rupprecht_2020,qiu2019roomtemperature}.
And, the control of valley coherence, which features a rotation of polarized light emission (referring to a rotation of coherent superposition of valley exciton states), has widely been reported by resorting to magnetic field~\cite{PhysRevLett.117.187401,PhysRevLett.117.077402}.
Further, both the direction and speed of coherent rotation can be manipulated  by tuning the dynamic phase of excitons hosted in distinct valleys~\cite{ye_sun_heinz_2017}.
However, despite substantial efforts, so far a fundamental microscopic model describing the dynamical evolution of valley (exciton) coherence in TMDCs, from which the coherence intensity and time---crucial quantities for valley manipulation--- can be directly inferred, is still not available, though is greatly desired from both fundamental physics and experimental application points of view.

Here, we develop a model for valley exciton dynamics in monolayer WS$_2$ subjected to magnetic field, by incorporating both intra- and intervalley scatterings associated with the exciton recombination, electron-hole (e-h) exchange and exciton-phonon (ex-ph) interactions, to master equation.
Then, we determine how the valley coherence evolves with time, and further unveil the potential means of how to enhance it in practice.
In addition to the LP excitation, which is widely adopted in experiments for coherent valley control~\cite{PhysRevLett.117.187401,PhysRevLett.117.077402},
we also demonstrate an emerging valley coherence of dark excitons even in the absence of initial coherent state.
To proceed, we first introduce our theoretical framework.

\begin{figure}
\includegraphics[width=1\linewidth]{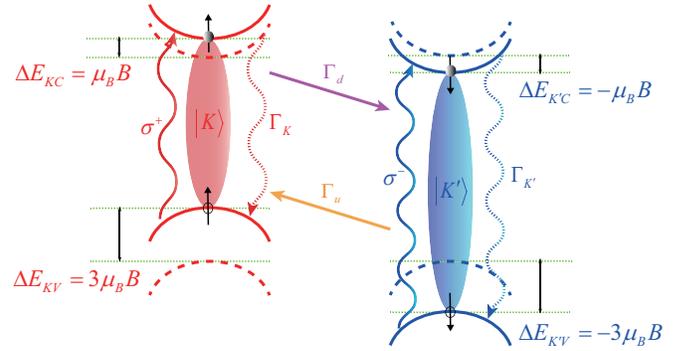}% Here is how to import EPS art
  \caption{Illustration of the spin-valley locked band structure in monolayer WS$_2$, with the magnetic field $B=0$ (dashed curves) and $B>0$ (solid curves).
  The red (blue) represents spin up (spin down) states in the $K$ ($K'$) valley,  $\sigma^{+}$ ($\sigma^{-}$) denotes  CP excitation,
  $\Delta E_{K(K')C(V)}$ stands for valley Zeeman shift for the conduction (valence) band,  and $\Gamma_{K(K')}$ is  the exciton recombination rate. The purple (orange) arrow denotes the intervalley scattering of excitons from a lower (higher) valley in energy to a higher (lower) one due to the e-h exchange and ex-ph interactions with the rate $\Gamma_{d(l)}$.}
\label{figure1}
\end{figure}

\textit{Intravalley scattering: exciton recombination.}---
The exciton recombination is one important process leading to valley decoherence.
Both the magnetic-field induced valley Zeeman shift (Fig.~\ref{figure1}) and the thermal (temperature) effect triggered band-gap ($E_{\rm g}$) shrinking affect the exciton energy~\cite{PhysRevLett.121.057402} and so the exciton recombination rate.
For the valley Zeeman shift, it comprises three distinct contributions from the spin, valley, and transition-metal atomic magnetic moments, while the magnetic response for the energy of \emph{intravalley} excitons is fully attributed to the atomic contribution~\cite{Koperski_2018,aivazianmagnetic2015}.
Among them, the valley magnetic moment does not affect the excitonic resonance of excitons within a given valley (i.e., intravalley exciton) that we consider, allowing us to ignore the valley contribution.
For the conduction band, the atomic contribution ($d_{z^2}$) is zero, and thus only the spin magnetic moment matters, with its contribution $\Delta_s^c=\mu_BB$, where $\mu_B$ is the Bohr magneton.
Regarding the valence band, both atomic ($d_{x^2-y^2}$ and $d_{xy}$) and spin magnetic moments contribute to the valley Zeeman shifts, with contributions $\Delta_a^v=2\mu_BB$ and $\Delta_s^v=\mu_BB$, respectively, leading to a total shift of $3\mu_BB$ (see Fig.~\ref{figure1}).
Note that for the bright exciton, this treatment of valley Zeeman shift with three contributions~\cite{Koperski_2018,aivazianmagnetic2015,PhysRevB.98.195302,PhysRevB.97.081109,Koperski_2019}, is consistent with recent \emph{ab initio} calculations, which deal with the valley (\textit{intercellular}) and atomic (\textit{intracellular}) contributions in a unified way associated with the orbital angular momentum~\cite{Faria2022,PhysRevLett.126.067403,PhysRevLett.124.226402,PhysRevB.101.235408}. However, when consider dark excitonic states or deal with van der Waals heterostructures of TMDCs, it is better to resort to more rigorous consideration treating the valley and atomic contributions in a unified way~\cite{Faria2022}.
Regarding the thermal effect, which causes a reduction of $E_{\rm g}$ primarily due to the electron-phonon interaction~\cite{PhysRevLett.121.057402}, we adopt a fit with the Varshni equation $\Delta E_{\rm g}=\alpha T^2/(T+\beta)$~\cite{VARSHNI1967149}, with $T$ the temperature, $\alpha\approx4.45\times10^{-4}~\rm{eV/K}$ a material relevant constant~\cite{PhysRevLett.121.057402} and $\beta\approx247.5~\rm{K}$ related to the Debye temperature~\cite{Helmrich_2018}.
Then, the exciton energy reads $E_{\xi}=E_{\rm g}-E_{\rm b}+{\uplambda_c}/{2}-{\uplambda_v}/{2}-2\xi \mu_BB$~\cite{2019}, with $E_{\rm b}$ the exciton binding energy, $\uplambda_{c(v)}$ the spin-orbit splitting in the conduction (valence) band, and $\xi=\pm1$ the valley indices $K$ and $K'$.
Note that the last term in $E_{\xi}$ describes the valley Zeeman shift. With all these considerations, the exciton recombination time is, $\uptau_\xi(T)=1/\Gamma_{\xi}(T)=({3Mc^2k_{B}T}/{2E_{\xi}^2})\uptau(0)$~\cite{doi:10.1021/nl503799t},
with $\uptau(0)=0.19~\rm{ps}$ the recombination time near $T=0$~\cite{doi:10.1021/nl503799t}, $M$ the exciton mass and $c$ the speed of light.

%\subsection{Exciton transfer with intervalley scattering}
\textit{Intervalley scattering: e-h exchange and ex-ph interactions.}---
The e-h exchange and ex-ph interactions in general dominate intervalley scatterings, providing a unique channel for transfer of excitons from one valley to the other~\cite{nanolett.8b01484,acs.jpcc.2c05113,zeng2012valley,PhysRevB.87.115418,PhysRevLett.127.157403,kioseoglou2016,carvalho2017,PhysRevA.103.043713}.
For the e-h exchange interaction, its magnitude essentially scales linearly with the center-of-mass wave vector $\mathbf{k}$ of excitons, and its direction depends on the orientation of $\mathbf{k}$~\cite{PhysRevB.107.035404,PhysRevLett.120.046402,PhysRevLett.121.057403},
and hence it provides an \emph{effective} in-plane magnetic field driving the precession of valley pseudospin with different frequencies~\cite{PhysRevB.96.125303,PhysRevB.104.L121408,PhysRevB.89.205303}.
Due to various scatterings, e.g., with phonons, other excitons and defects, the exchange interaction, which depends on $\mathbf{k}$, may become random~\cite{nanolett.7b03953}.
This is similar to the D'ykonov-Perel (DP) spin relaxation~\cite{fabian2007,PhysRevB.90.245302,PhysRevB.103.165304}, induced by the momentum-dependent spin-orbit field, for two-dimensional electron gases in the diffusive regime.
The incoherent intervalley transfer caused by the random exchange interaction manifests as a statistical average scattering time $\uptau_{\rm{v0}}$.
Further, in the presence of external magnetic field, the expectation value of valley pseudospin depends on the combined contributions of the in-plane and out-of-plane components, indicating that the influences of exchange interaction can be effectively tuned by the valley Zeeman splitting $\Delta E=E_{K'}-E_K$~\cite{PhysRevB.107.035404}.
Thus, the magnetic-field mediated intervalley scattering rate reads $\Gamma_{\rm{v}}=1/\uptau_{\rm{v0}}\times F(\Delta E)$, where $F(\Delta E)=\Gamma^2/(\Gamma^2+\Delta E^2)$~\cite{nanolett.8b01484,acs.jpcc.2c05113}, with $\uptau_{\rm{v0}}=50$ ps the zero field scattering time and $\Gamma=0.1~\rm{meV}$ the width parameter associated with exciton momentum relaxation time~\cite{nanolett.8b01484}, i.e., $\Gamma=\hbar/\uptau_p$.
Note that the magnitude of magnetic field at which the e-h exchange interaction is suppressed can be estimated by analyzing the momentum relaxation time $\uptau_p$~\cite{PhysRevB.50.14246,PhysRevB.47.15776}.

Regarding the ex-ph interaction, since the excitons in both $K$ and $K'$ valleys have zero center-of-mass momentum, two $K$-point phonons are needed to ensure the momentum conservation for the transfer process.
In such a situation, the intervalley scattering rate is proportional to the phonon occupation number, i.e., $\Gamma_{\rm{ph}}\propto\exp(-\langle\hbar\omega_{\rm{ph}}\rangle/k_BT)$~\cite{zeng2012valley}, with $\langle\hbar\omega_{\rm{ph}}\rangle=12~\rm{meV}$ the acoustic phonon energy near the $K$-point~\cite{zeng2012valley,PhysRevB.87.115418,PhysRevLett.127.157403,kioseoglou2016,carvalho2017}, closing to the acoustic phonon energy reported in TMDCs~\cite{PhysRevB.44.3955,zeng2012valley,Helmrich_2018}.
Further, to capture the \emph{asymmetry} of the phonon-related relaxation process caused by the lifting of valley degeneracy due to $B$ field, the intervalley scattering rates associated with ex-ph interaction are expressed as Miller form, with $\Gamma_{H}=\Gamma_{\rm{ph}}=1/50$ ps$^{-1}$ and $\Gamma_{L}=\Gamma_{\rm{ph}}\exp(-\Delta E/k_BT)$~\cite{PhysRev.120.745,doi:10.1063/1.5112823,PhysRevApplied.18.064022,Baranowski2017}.
This ensure that excitonic scatterings from the valley of higher energy to the one of lower energy require emission of an additional phonon ($\Gamma_H$), whereas absorption of a phonon occurs in the opposite process ($\Gamma_L$), which is mediated by the Boltzmann factor and reduced to $\Gamma_{H}$ at $B=0$ (symmetric intervalley scattering).
Combining two part contributions from e-h exchange and ex-ph interactions, the total intervalley scattering rates can be written as $\Gamma_{u,d}=\Gamma_{\rm{v}}+\Gamma_{H,L}$.

\textit{Pure dephasing of valley exciton: additional coherence decay.}---
In addition to the exciton recombination, pure dephasing is also a significant source of coherence loss.
Considering the Maialle-Silva-Sham (MSS) mechanism, an in-depth understanding of processes responsible for valley pure dephasing is exchange interaction~\cite{PhysRevB.50.10868,hao_moody2016}.
Further, optical 2D coherent spectroscopy reveals another pure dephasing pathway, by elevating temperature~\cite{moody2015intrinsic,hao_moody2016}, which is analogous to the exciton dephasing in semiconductor quantum wells caused by the acoustic phonon scattering~\cite{PhysRevB.34.9027}.
Hence the pure dephasing rate is expressed as $\upgamma=\upgamma_0+\upgamma_{\rm{eh}}^0F(\Delta E)+\upgamma_{\rm{ph}}^0T$,
where $\upgamma_0$ is the residual exciton dephasing rate in the absence of e-h exchange and ex-ph interactions~\cite{moody2015intrinsic}.
$\upgamma_{\rm{eh}}^0=2~\rm{ps}^{-1}$ is the dephasing rate related to exchange interaction at zero extra field~\cite{hao_moody2016}, and factor $F(\Delta E)$ stands for the suppression by the magnetic field~\cite{nanolett.8b01484,acs.jpcc.2c05113}.
$\upgamma_{\rm{ph}}^0=1/12~\rm{ps}^{-1}/\rm{K}$ denotes the ex-ph coupling strength~\cite{moody2015intrinsic}.
The values of these parameters can be further analysed experimentally from the homogeneous linewidth~\cite{selig2016excitonic,moody2015intrinsic}.
We exhibit the numerical relationship between the rate $\upgamma_{\rm{eh}}$ ($\upgamma_{\rm{ph}}$) and the magnetic field (temperature) in Fig.~S1 of Supplemental Material (SM).
When $B>5$ T, the exchange interaction is completely suppressed by the vertical magnetic field, so that the pure dephasing rate $\upgamma_{\rm{eh}}$ trends to zero.
In contrast, the rate $\upgamma_{\rm{ph}}$ increases linearly with elevated temperature that highlights the decisive role of phonon scattering at high temperatures.

\textit{Valley dynamical evolution: master equation.}---
We incorporate the intra- and intervalley scatterings into the master equation of Lindblad form, and employ the density matrix to illustrate evolutions of the valley coherence and exciton population.
We adopt the basis set $\{|K\rangle,|K'\rangle,|0\rangle\}$ characterizing the exciton states in the $K$ ($|K\rangle$) and $K'$ ($|K'\rangle$) valleys  and the ground state ($|0\rangle$).
Also, note that the intervalley transfer of excitons is a process of \emph{balancing} populations between the two valleys, which is described by two incoherent rate equations~\cite{qiu2019roomtemperature}, see Eq.~(S5) of SM, while the exciton recombination refers to a one-way flow of valley information to the environment~\cite{PhysRevLett.103.210401,PhysRevB.101.174302}.
Then, the Hamiltonian of the overall system reads, $H=H_0+H_{\text{E}}+H_\text{I}$, comprising contributions from the valley exciton ($H_0$)~\cite{PhysRevLett.123.096803}, environment ($H_{\text{E}}$) and interaction between valley and environment ($H_{\text{I}}$),
\begin{equation}
\label{eq1}
\begin{split}
H_0&=\sum_{\xi}E_{\xi}c_{\xi}^{\dag}c_{\xi},~\xi=K,K',\\
H_{\text{E}}&=\hbar(\omega_La_L^{\dag}a_L+\omega_Ra_R^{\dag}a_R),\\
H_{\text{I}}&=g_K\sigma_{K-}a_R^{\dag}+g_{K'}\sigma_{K'-}a_L^{\dag}+\rm{H.c.},
\end{split}
\end{equation}
where $c_{\xi}^{\dag}(c_{\xi})$ stands for the exciton creation (annihilation) operator of the $\xi$ ($=K,K'$) valley.
$a_L^{\dag}(a_L)$ and $a_R^{\dag}(a_R)$ denote the creation (annihilation) operators of the left ($L$) and right ($R$) circularly polarized photons with the characteristic frequencies $\omega_L$ and $\omega_R$.
Constant $g_\xi$ is the coupling strength between the exciton in $\xi$ valley with the corresponding circularly photon mode.
Also, we have defined $\sigma_{\xi-}=|0\rangle\langle\xi|$ as the lowering operator and $\rm{H.c.}$ denotes the Hermitian conjugate, indicating the exciton recombination are from individual $K$ and $K'$ valleys, with emitting right and left circularly polarized photons, respectively.
Thus, the dynamical evolution of exciton valley coherence is described as,
\begin{equation}
\label{eq2}
\begin{split}
\frac{d}{dt}\rho_t=L(\rho_t)=L_{0}(\rho_t)+L_{\rm{f}}(\rho_t)+L_{\rm{r}}(\rho_t)+L_{\rm{p}}(\rho_t),
\end{split}
\end{equation}
where $L_0(\rho_t)$ denotes the unitary evolution determined by the exciton Hamiltonian $H_0$, and $L_{\rm{f}}(\rho_t)$ means the exciton intervalley transfer between two valleys.
$L_{\rm{r}}(\rho_t)$ [$L_{\rm{p}}(\rho_t)$] describes the exciton recombination (pure dephasing) process, referring to the rate $\Gamma_\xi$ ($\upgamma$).
For more details about the operator $L(\rho_t)$, see the SM.
As the valley coherence strongly dependents on the remnant excitonic populations, we define the coherence intensity characterizing
the degree of valley coherence by employing the $l_1$-norm-based coherence measure~\cite{PhysRevLett.113.140401,Lan_2022},
which in the basis set of $\rho_t=(\rho^{K},\rho^{K'},\rho^{K'K},\rho^{KK'},\rho_0)^{T}$ reads,
\begin{equation}
\label{eq3}
\begin{split}
\mathcal{C}(\rho)=\sum_{i\not=j}|\rho^{ij}|=|\rho^{KK'}|+|\rho^{K'K}|,
\end{split}
\end{equation}
with $0\leq\mathcal{C}(\rho)\leq1$.
And, the coherence time $\uptau_{\mathcal{C}}$ can be defined as a time over which $\mathcal{C}(\rho)$ essentially vanishes~\cite{footenote-tauc}, characterizing the duration of coherence.

\begin{figure}[h!]
\includegraphics[width=0.9\linewidth]{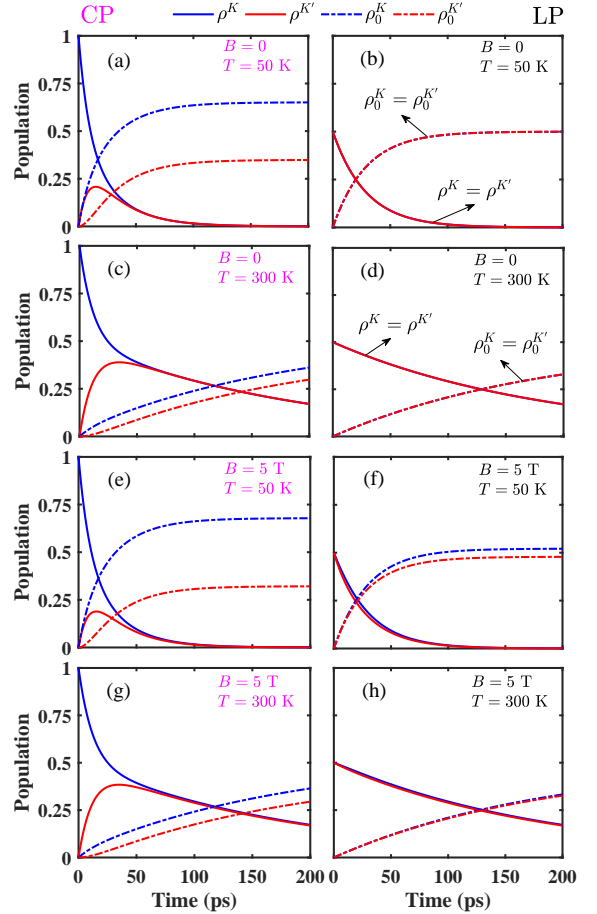}% Here is how to import EPS art
\caption{Time evolutions of exciton populations for different magnetic fields and temperatures in the cases of CP (left panels) and LP (right panels) excitations.
The density matrix element $\rho^{K}$ ($\rho^{K'}$) stands for the remnant exciton population after recombination in the $K$ ($K'$) valleys, while $\rho_0^{K}$ ($\rho_0^{K'}$) refers to the population of exciton undergoing recombination.}
\label{figure2}
\end{figure}

\textit{Time evolutions of density matrix elements.}---
For better understanding valley dynamics, we first look into how the exciton populations represented by the density matrix elements $\rho^{K(K')}$ and $\rho_0^{K(K')}$, evolve with time in the presence of both intra- and intervalley scatterings (Fig.~\ref{figure2}).
We focus on optical pumping of both CP and LP excitations, which gives rise to the \emph{initial} states of $|\uppsi_0\rangle=|K\rangle$ (e.g., for $\sigma^+$ excitation) and $|\uppsi_0\rangle=1/\sqrt{2}(|K\rangle+|K'\rangle)$, respectively, as a result of valley selective transition rule.

For the CP excitation (left panels in Fig.~\ref{figure2}), as the excitons are initially generated in the $K$ valley with $|\uppsi_0\rangle=|K\rangle$,
we find that $\rho^{K'}$ first increases and then decreases after attaining its maximal value, due to a combined effect of \emph{intervalley} excitonic transfer and \emph{intravalley} exciton recombination, in contrast to $\rho^{K}$, which consistently decreases, cf. $\rho^{K}$ and $\rho^{K'}$ in Fig.~\ref{figure2}(a).
Now we examine the effect of temperature and magnetic field, both of which are found  playing an important role in valley dynamics.
As temperature increases, on the one hand, the process of exciton recombination is quenched, giving rise to considerable increment of $\uptau_{K(K')}$; on the other hand, phonon assisted intervalley scattering becomes more pronounced, leading to more balanced populations between $K$ and $K'$ valleys, cf. Figs.~\ref{figure2}(a) and \ref{figure2}(c).
As opposed to  $\rho^{K,K'}$, $\rho_0^{K,K'}$ describing the population of recombined excitons, consistently increase with time, as is expected.
Regarding the $B$ field, which lifts the valley degeneracy and introduces the asymmetry of intervalley scatterings (i.e., unbalanced $\Gamma_H$ and $\Gamma_L$), thus refraining the exciton transfer from the $K$ to $K'$ valleys, resulting in an overall reduction of exciton population in the $K'$ valley, cf. Figs.~\ref{figure2}(a) and \ref{figure2}(e).

As for the LP excitation (right panels in  Fig.~\ref{figure2}), because of an \emph{initial} coherent superposition of excitonic states in the two valleys with the same population, the dynamical evolution of $\rho^{K}$ and $\rho^{K'}$ perfectly matches at zero $B$ field, independent of
temperature, see Figs.~\ref{figure2}(b) and \ref{figure2}(d) for $T=50$ and 300 K, respectively.
Similarly, the populations for recombined excitons between the $K$ and $K'$ valleys are also locked to be equal as time evolves, cf. $\rho_0^{K}$ and $\rho_0^{K'}$, though $\rho^{K(K')}$ and $\rho_0^{K(K')}$ exhibit opposite dynamic behavior.
Further, even with the presence of valley Zeeman splitting, the distinction of excitonic populations between the two valleys is also greatly quenched [Figs.~\ref{figure2}(f) and \ref{figure2}(h)], as compared to the case of CP excitation, cf. left and right panels.

These dynamical features of density matrix elements are helpful in understanding valley dynamics.
For further illustrating the coherence behaviors, below we focus on the LP excitation, which is widely employed in experiments
for coherent valley control~\cite{PhysRevLett.117.187401,PhysRevLett.117.077402}.

\begin{figure}
\includegraphics[width=\linewidth]{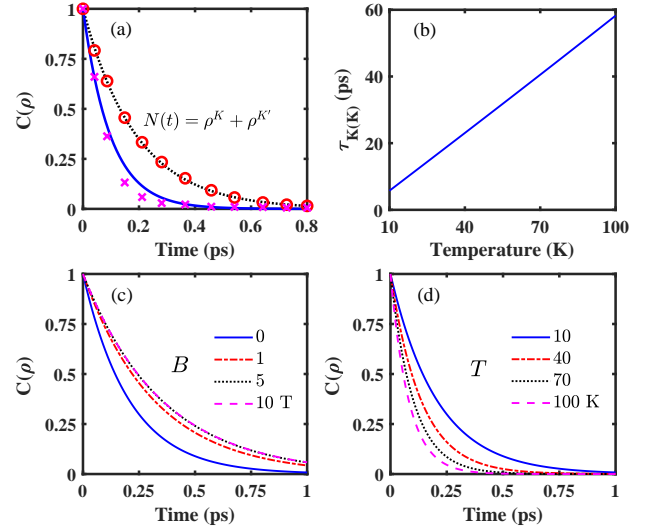}% Here is how to import EPS art
\caption{(a) Time evolution of valley coherence (solid line; blue) and remnant exciton population (dashed line; black).
  The markers (crosses and circles) refer to the corresponding  experimental data of Ref.~[\onlinecite{hao_moody2016}].
(b) The exciton recombination lifetime as function of temperature~\cite{doi:10.1021/nl503799t,C5NR00383K}.
(c), (d) Valley coherence dynamics at different magnetic fields ($T=10~\rm{K}$) (c) and temperatures ($B=0$) (d).}
\label{figure3}
\end{figure}

\textit{LP excitation: magnetic-field enhanced coherence.}---
The main channels leading to the decay of valley coherence comprises the intravalley exciton recombination ($\Gamma_{K(K')}$) and the pure dephasing process ($\upgamma$).
The former results in reduction of excitonic populations (Fig.~\ref{figure2}), and the latter mainly originates from the e-h exchange and ex-ph interactions.
For verifying the accuracy of our model, we first adopt experimental parameters in the work by Hao \emph{et al}.~\cite{hao_moody2016},
where the decoherence rate $\upgamma_v=\Gamma_{K}+\upgamma=10.2~\rm{ps}^{-1}$ with $\Gamma_{K}=5.26$ and $\upgamma=4.94~\rm{ps}^{-1}$~\cite{moody2015intrinsic,hao_moody2016}.
We reveal that the remnant exciton population, $N=\rho^{K}+\rho^{K'}$, tends to vanish at around $800~\rm{fs}$ [dotted line in Fig.~\ref{figure3}(a)], while the coherence time $\uptau_{\mathcal{C}}$ attains about $400$ {fs} [solid line in Fig.~\ref{figure3}(a)], in great agreement with experimental data [markers in Fig.~\ref{figure3}(a)].
It is noteworthy that experimental reports of exciton lifetime in monolayer WS$_2$ usually ranges from tens to even hundreds of ps~\cite{PhysRevB.90.075413,C5NR00383K,PhysRevLett.112.047401,liu_wang2020,PhysRevLett.123.067401}.
Considering that the relaxation of excitonic population may provide an upper bound for valley coherence,
below we consider the exciton lifetime ranging from 5 to 60 ps as the temperature varies from 10 to 100 K [Fig.~\ref{figure3}(b)] ~\cite{doi:10.1021/nl503799t,C5NR00383K}, to explore potential means of enhancing valley coherence.

We first examine the magnetic response of valley coherence dynamics [Fig.~\ref{figure3}(c)].
It is found that the maximal coherence intensity occurs \emph{initially}, following that the initial state itself being a coherent superposition of valley excitons.
Notably, when $B<5~{\rm T}$,  we reveal magnetic-field bolstered valley coherence in both its intensity and time,
which arises from the quenching of the pure dephasing channel induced by the e-h exchange
interaction as a result of lifting of valley degeneracy in the presence of $B$ field.
However, for $B>5$ T we observe that the coherence dynamics starts to remain essentially unaltered as $B$ varies, cf. dynamics behaviors at $B=5$ and $10~\rm{T}$.
This is because the pure dephasing induced by e-h exchange interaction becomes entirely suppressed at $B=5$ T [see Fig.~S1(a)] and the other decoherence channels is basically unaffected by the $B$ field.
Note that the coherence time at $T=10$ K even attains $\uptau_{\mathcal{C}}\sim1$ ps, facilitating complete control of qubits based on the valley pseudospin.

Now we turn to the temperature related thermal effects, which essentially have two \emph{compensating} consequences on valley coherence.
On the one hand, the ex-ph interaction becomes more pronounced at escalated temperature, in favor of decay of coherence; on the other hand,
a higher temperature implies a longer exciton recombination time, thus suppressing the valley decoherence.
However, the \emph{overall} effect is that the ex-ph interaction with increasing temperature gradually dominates over the other decoherence channels due to, i.e., the e-h exchange interaction or the exciton recombination [cf. Fig.~S1(b) and Fig.~\ref{figure3}(b)].
Thus, a shrinking of valley coherence at escalated temperature follows [Fig.~\ref{figure3}(d)].

Despite the detrimental effect of temperature on valley coherence, we should emphasize the suppression of e-h exchange interaction by magnetic field
provides a reliable route to enhance coherence intensity and enlarge the coherence time in a temperature range of $T=10-120~\rm{K}$.
Also, the robustness of the coherence against $B$ field when $B>5~\rm{T}$ [Fig.~\ref{figure3}(c)] also holds in a broad temperature range (Fig.~S2 of SM), facilitating coherent manipulation of the valley pseudospin.
For practical application, a possible means of boosting coherence generation is to resort to the exciton-cavity coupling~\cite{qiu2019roomtemperature},
which may open an additional channel of coherence transfer from photons to valley excitons.

\textit{Emerging valley coherence of dark excitons: with no initial coherence.}---
Under $\sigma^+$ CP excitation, only excitons in the $K$ valley is generated and thus there is no initial coherence.
In this case, the incoherent intervalley transfer prevents exciton dynamics from producing valley coherence.
In contrast, for dark excitons, the exchange interaction is a short-range local field effect, thus it weakly depends on center-of-mass momentum and can be treated as a constant with definite phase.
In the basis $\{|K_d\rangle,|K'_d\rangle\}$, the system Hamiltonian of dark exciton has the form~\cite{PhysRevLett.123.096803}
\begin{equation}
\label{eq4}
\begin{split}
H_0^d=
\left(                 %左括号
 \begin{array}{ccc}   %该矩阵一共4列，每一列都居中放置
  E_{K}^d+\delta&\delta\\
  \delta&E_{K'}^d+\delta
  \end{array}
\right),
\end{split}
\end{equation}
with $E_{\xi}^d$ the dark exciton energy, and $\delta$ the short-range exchange interaction~\cite{PhysRevLett.123.096803,PhysRevB.96.155423}, that \emph{mixes} different valley states together.
Also, the dark excitons have a longer recombination lifetime than bright state, due to the weak coupling with photons of out-of-plane linearly polarization.
These features underline a potential coherent control based on dark excitons, even without initial coherence, as we discuss next.

For completeness, we consider two initial states of $|\psi_0^d\rangle=1/\sqrt{2}(|K_d\rangle+|K'_d\rangle)$ (with initial coherence) and $|\psi_0^d\rangle=|K_d\rangle$ (without initial coherence).
The relevant parameters adopted in the simulation are $\delta=0.6$ meV~\cite{PhysRevLett.123.096803,PhysRevB.96.155423}, recombination lifetime $\uptau_\xi^d=110$ ps~\cite{PhysRevB.96.155423}, and pure dephasing rate $\upgamma^d=2.6$ ps$^{-1}$ at $T=10$ K, with the super/subscript $d$ denoting the dark.
For detail derivation of valley coherent dynamics of dark excitons, see the SM.

\begin{figure}
\includegraphics[width=\linewidth]{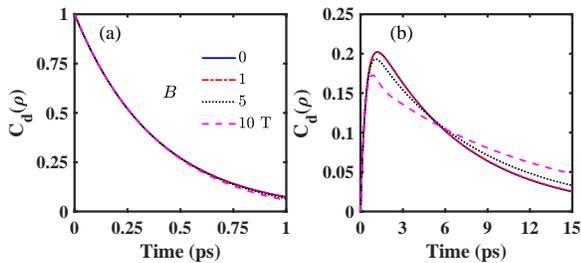}% Here is how to import EPS art
\caption{Valley coherence dynamics of dark excitons at different magnetic fields ($T$=10 K), for initial coherent state (a) and initial incoherent state (b).}
\label{figure4}
\end{figure}
Figure~\ref{figure4}(a) shows the magnetic response of coherence behaviors with initial coherence.
In this case, we find that despite the long recombination lifetime of dark exciton at low temperature, its coherent evolution essentially coincides with that of the bright state, cf. Fig.~\ref{figure3}(c) and Fig.~\ref{figure4}(a).
This is because valley coherence mainly depends on the pure dephasing in the presence of initial coherent state.
In addition, a constant exchange interaction avoids triggering the MSS mechanism, which means that the $B$ field has negligible impact on the pure dephasing path.
Hence the coherence intensity and time of dark excitons are not sensitive to the $B$ field [see Fig.~\ref{figure4}(a)].
Regarding the case of without initial coherence, it is shown that the valley coherence can be generated through the \emph{mixture} of excitonic states of two valleys by exchange interaction [Fig.~\ref{figure4}(b)].
In contrast to the case of with initial coherence, this emerging coherence depends not only on the generation channel, but also on the decay channel (i.e., decoherence), thus leading to a long coherence time ($\uptau_{\mathcal{C}}^d\sim15$ ps), though with quenched coherence intensity.
Also, the \emph{mixture} effect of exchange interaction can be tuned by the valley splitting, which manifests as a marked dependence of this emerging coherence on $B$ field.

We should emphasize that, in general, as dark excitons are generated by the intravalley scattering from bright excitons~\cite{darkf,2019,PhysRevLett.123.096803,Baranowski2017}, a more precise description of coherent valley dynamics requires to take into account the bright and dark states simultaneously.
In particular, when the magnetic field has an in-plane component, it even mixes the bright and dark states.
More work is needed to explore these interesting possibilities.

%\section{Conclusions}
%\label{sec:conc}
\textit{Conclusions.}---
The artificial manipulation of valley pseudospin requires a sufficiently advanced coherence quality.
We have developed a microscopic model in monolayer WS$_2$ involving both intra- and intervalley scatterings, and unveiled magnetically tunable exciton valley coherence mediated by the e-h exchange and ex-ph interactions.
For the LP excitation, which is accompanied with an \emph{initial} coherence, our determined valley dynamics manifests the coherence decay being faster than the exciton population relaxation, and agrees with experimental data by Hao \emph{et al.} [\href{https://dx.doi.org/10.1038/nphys3674}{Nat. Phys. 12, 677 (2016)}].
Further, we find that magnetic field may quench the e-h exchange induced pure dephasing, allowing to \emph{magnetically} regulate valley coherence.
In particular, at low temperatures for which the ex-ph interaction is weak, the coherence time attains $1$ ps, a significant improvement over previously reported values in the sub-ps scale~\cite{hao_moody2016,PhysRevLett.116.127402,moody2015intrinsic}.
For practical considerations, we should emphasize that the density of exciton gas may also affect the coherence dynamics.
A higher density will enhance the strength of exchange interaction~\cite{nanolett.7b03953,moody2015intrinsic,PhysRevB.103.045426}, which requires a stronger $B$ field ($\textgreater5$ T) to suppress the pure dephasing path.
Correspondingly, a higher $B$ field is needed for the valley coherence to exhibit robustness.
As a remark, the coherence time is expected to be further enlarged through external means, e.g., exciton-cavity coupling~\cite{qiu2019roomtemperature}, electron doping~\cite{yangsinitsyn2015}, and enhanced dielectric screening~\cite{gupta2021}.
For the dark excitons, we observe an emerging valley coherence even in the absence of initial coherent state.
And, this emerging coherence has long coherence time ($\sim15$ ps), though with quenched coherence intensity.
Our work provides an insight into \emph{tunable} valley coherence and coherent valley control based on dark excitons.

We thank Paulo E. Faria Junior and Gerson J. Ferreira for valuable discussions.
This work was supported by the National Natural Science Foundation of China (Nos. 11974212, 12274256 and 11874236), and the Major Basic Program of Natural Science Foundation of Shandong Province (Grant No. ZR2021ZD01).

\end{CJK}
%\bibliography{myref}

\begin{thebibliography}{79}
\bibitem [{foo()}]{footenote-tauc}%
  \BibitemOpen
  \href@noop {} {}\bibinfo {note} {The coherence time is also widely defined as
  a time over which the coherence intensity decays to $e^{-1}$ of its initial
  value.}\BibitemShut {Stop}%
\end{thebibliography}

\end{document}